\documentclass[twocolumn,showpacs,preprintnumbers,aps]{revtex4}
\usepackage{graphicx}

\begin{document}

\title{Magnetization plateaux, Haldane-like gap, string order and hidden
symmetry in a spin-1/2 tetrameric Heisenberg antiferromagnetic chain}
\author{Shou-Shu Gong and Gang Su$^{{\ast }}$}
\affiliation{College of Physical Sciences, Graduate University of Chinese Academy of
Sciences, P. O. Box 4588, Beijing 100049, China}

\begin{abstract}
The ground-state properties of a spin $S=1/2$ tetrameric Heisenberg
antiferromagnetic chain with alternating couplings AF$_{1}$-AF$_{2}$-AF$_{1}$%
-F (AF and F denote antiferromagnetic and ferromagnetic couplings,
respectively) are studied by means of the density matrix renormalization
group method. Two plateaux of magnetization $m$ are found at $m=0$ and $1/4$%
. It is shown that in such a spin-$1/2$ AF system, there is a gap from the
singlet ground state to the triplet excited states in the absence of a
magnetic field. The spin-spin correlation function decays exponentially, and
the gapped states have a nonvanishing string order, which measures a hidden
symmetry in the system. By a dual transformation, the string order is
transformed into a ferromagnetic order and the hidden symmetry is unveiled
to be a $Z_{2}\times Z_{2}$ discrete symmetry, which is fully broken in the
gapped states. This half-integer spin Heisenberg AF chain is in a
Haldane-like phase, suggesting that the present findings extend the
substance of Haldane's scenario. A valence-bond-solid state picture is also
proposed for the gapped states.
\end{abstract}

\maketitle

\section{Introduction}

Even though many models have been extensively studied experimentally and
theoretically, one-dimensional quantum spin systems are still an attractive
field in low-dimensional quantum magnetism. One of motivations is from
Haldane's conjecture \cite{Haldane}. Haldane proposed that an isotropic
Heisenberg antiferromagnetic chain (HAFC) with integer spin has a finite gap
from the singlet ground state to the triplet excited states, and the
spin-spin correlation function decays exponentially; while the HAFC with
half-integer spin has a gapless spectrum and a correlation function with
power-law decay. Although Haldane's conjecture has been confirmed
experimentally and numerically in many systems, there is not a rigorous
proof till now. By means of the valence bond states, a rigorous disordered
ground state for the biquadratic Heisenberg Hamiltonian with spin $S=1$ was
proposed by Affleck, Kennedy, Lieb and Tasaki (AKLT) \cite{AKLT}. This AKLT
model was shown to have a spin gap from the singlet ground state to the
triplet excited states, an exponentially decaying correlation function, and
a nonlocal topological order which is different from the long-range order of
spin-spin correlation, thus confirming the Haldane's conjecture based on a
specific model. This nonlocal string order, which is regarded as a hidden
antiferromagnetic N\'{e}el order, was found by Den Nijs and Rommelse \cite%
{NR} in the spin-$1$ HAFC. Kennedy and Tasaki \cite{KT} introduced a
nonlocal unitary transformation to reveal the hidden symmetry for this
string order, which was found to be a discrete $Z_{2}\times Z_{2}$ symmetry.
The Haldane phase of integer-spin HAFC is thus characterized by the complete
breaking of the $Z_{2}\times Z_{2}$ symmetry \cite{KT}.

Besides the isotropic HAFC with integer spin, Haldane phase has also been
found in other spin systems, like some $S=1/2$ spin ladders \cite%
{Hida1,WNT,BR} and spin-$1/2$ ferromagnetic-antiferromagnetic (F-AF)
alternating Heisenberg chain \cite{Hida2}. In these systems, the gap and the
string order vary monotonically with the ratio between F and AF interactions
($J_{F}/J_{AF}$) and when $J_{F}\rightarrow -\infty $, recover the values
obtained for the spin-$1$ HAFC. The weak F coupling phase belongs to the
same phase as the strong F coupling case. When the F couplings dominate the
AF ones, the dimers of two spins coupled by F interactions behave as spin-$1$%
, and the systems reduce to the spin-$1$ HAFC. The $Z_{2}\times Z_{2}$
symmetry is also fully broken in these systems, which indicates that the
systems belong to the Haldane phase.

In this paper, we are concerned about the existence of a Haldane-like phase
in an AF chain with half-integer spin, which, unlike the spin-$1/2$ F-AF
alternating chain, could not be reduced to a HAFC with integer spin in any
circumstance. The spin-$1/2$ trimerized F-F-AF Heisenberg chain might be a
choice, but its spectrum is gapless \cite{Gu}. Here we consider a spin $%
S=1/2 $ tetrameric HAFC with alternating couplings AF$_{1}$-AF$_{2}$-AF$_{1}$%
-F. As far as we know, there has been no report on any tetrameric
antiferromagnet yet, and only has a tetrameric ferrimagnetic Heisenberg
chain compound Cu(3-Clpy)$_{2}$(N$_{3}$)$_{2}$ (F-F-AF-AF) been widely
studied \cite{HMK,HNMK,EVF,Yamamoto,NY,LSS}. By means of the density matrix
renormalization group (DMRG) method and dual transformation, we have found
that this spin-1/2 tetrameric system is in a gapped phase, most properties
of which are compared with the features of the Haldane phase. Most
importantly, the phase we found has a string order and a hidden $Z_{2}\times
Z_{2}$ symmetry, which is also fully broken in this system. This alternating
tetrameric HAFC with spin-$1/2$ is in a Haldane-like gapped phase. In this
sense, our findings extend the substance of Haldane's scenario, which
implies that the spin gap can also appear in certain half-integer spin HAFCs.

The rest of this paper is organized as follows. In Sec. \uppercase%
\expandafter{\romannumeral2}, we shall introduce the model Hamiltonian, and
give a brief discussion on the isolated tetramer systems. In Sec. \uppercase%
\expandafter{\romannumeral3}, we shall present our DMRG results on the
magnetic properties of the system in a longitudinal magnetic field, as well
as the gap behaviors, spin-spin correlation function and string order in the
gapped ground states in absence of the external field. A dual transformation
is introduced to the model in Sec. \uppercase\expandafter{\romannumeral4} to
unveil the hidden symmetry. In Sec. \uppercase\expandafter{\romannumeral5},
we shall propose a valence-bond-solid state picture and a trial wavefunction
for the gapped states. Finally, a summary and discussion will be given in
Sec. \uppercase\expandafter{\romannumeral6}.

\section{Spin-$1/2$ tetrameric Heisenberg antiferromagnetic chain}

\begin{figure}[tbp]
\includegraphics[width=0.8\linewidth,clip]{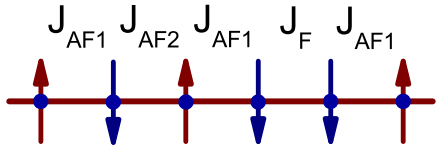}
\caption{(Color online) Schematic representation of the tetrameric
Heisenberg spin chain with alternating couplings described by Hamiltonian (%
\protect\ref{model Hamiltonian}). }
\label{spin configuration}
\end{figure}

The Hamiltonian of the isotropic tetrameric HAFC with alternating couplings
AF$_{1}$-AF$_{2}$-AF$_{1}$-F in a longitudinal magnetic field is given by 
\begin{eqnarray}
H &=&\sum\limits_{j}(J_{AF_{1}}\mathbf{S}_{4j-3}\cdot \mathbf{S}%
_{4j-2}+J_{AF_{2}}\mathbf{S}_{4j-2}\cdot \mathbf{S}_{4j-1}  \nonumber \\
&+&J_{AF_{1}}\mathbf{S}_{4j-1}\cdot \mathbf{S}_{4j}-J_{F}\mathbf{S}%
_{4j}\cdot \mathbf{S}_{4j+1})-h\sum\limits_{j}S_{j}^{z},
\label{model Hamiltonian}
\end{eqnarray}%
where $J_{AF_{1,2}}$ ($>0$) denote the different AF couplings, $J_{F}$ ($>0$%
) denotes the F coupling, and $h$ is the external magnetic field. We take $%
g\mu _{B}=1$ for convenience. The schematic spin configuration for the
tetrameric HAFC is depicted in Fig. \ref{spin configuration}. Here we would
like to remark that a spin-$5/2$ tetrameric HAFC compound, C$_{44}$H$_{36}$N$%
_{16}$Mn$_{2}$, with alternating couplings AF$_{1}$-AF$_{2}$-AF$_{1}$-F was
synthesized experimentally \cite{37Mn}. Since the DMRG calculations
on most physical properties of such a spin-$5/2$ tetrameric HAFC are
impermissible for our present computing capacity, and considering that the
spin-$1/2$ tetrameric HAFC is more fundamental and interesting, and can be
readily made comparisons to other Haldane spin systems, we opt to pay
attention to the tetrameric system with spin-$1/2$ in the following.
Meanwhile, for the reasons of comparison, some results on the spin-$5/2$
tetrameric HAFC will also be calculated.

Let us first discuss some cases of the present model. When $J_{AF_{1}}=0$,
the system is decoupled into isolated dimers with $J_{AF_{2}}$ or $J_{F}$
couplings. The dimers coupled by $J_{AF_{2}}$ form local singlets while
those coupled by $J_{F}$ form local triplets. Thus, the excitation of the
system from $S^{z}_{tot}=0$ to $S^{z}_{tot}=\pm1$ is gapless due to the
triplets.

When $J_{AF_{2}}=0$, the system is decoupled into isolated tetramers, which
becomes exactly soluble. In the absence of an external field, the ground
state of the tetramer is antiferromagnetic with $S_{tot}=0$, and the first
excited state is threefold degenerate with $S_{tot}=1$. The gap $\Delta$ is 
\begin{equation}
\Delta=\frac{1}{2}(\sqrt{4J_{AF_{1}}^{2}+2J_{AF_{1}}J_{F}+J_{F}^{2}}-J_{F}),
\label{gap_Jf}
\end{equation}
which increases monotonically with $J_{AF_{1}}$ but decreases with $J_{F}$.
When $J_{F}\gg J_{AF_{1}}$, the gap has an asymptotic form $\Delta\sim
J_{AF_{1}}(1+3J_{AF_{1}}/2J_{F})/2$ and equals $J_{AF_{1}}/2$ when $%
J_{F}\rightarrow\infty$.

In the limit $J_{F}=0$, the ground state of the tetramer is also
antiferromagnetic with $S_{tot}=0$, and the first excited state is threefold
degenerate with $S_{tot}=1$. The gap $\Delta$ from the ground state to the
first excited state is 
\begin{eqnarray}
\Delta&=&\frac{1}{2}(J_{AF_{1}}+\sqrt{%
4J_{AF_{1}}^{2}-2J_{AF_{1}}J_{AF_{2}}+J_{AF_{2}}^{2}}  \nonumber \\
&-&\sqrt{J_{AF_{1}}^{2}+J_{AF_{2}}^{2}}),  \label{gap_J2}
\end{eqnarray}
which enhances monotonically with $J_{AF_{1}}$ but decreases with $%
J_{AF_{2}} $. When $J_{AF_{2}}\gg J_{AF_{1}}$, the gap behaves as $%
\Delta\sim J_{AF_{1}}^{2}/2J_{AF_{2}}$ and vanishes when $%
J_{AF_{2}}\rightarrow\infty$.

In the large $J_{F}$ limit, the two spins coupled by $J_{F}$ form a spin-$1$%
, and the model reduces to a Heisenberg chain with alternating spin-($1$,$%
1/2 $,$1/2$): 
\begin{eqnarray}
H_{J_{F}\rightarrow\infty}&=&\sum_{j}(\frac{1}{2}J_{AF_{1}}\mathcal{S}%
_{3j-2}\cdot \mathbf{S}_{3j-1}+J_{AF_{2}}\mathbf{S}_{3j-1}\cdot \mathbf{S}%
_{3j}  \nonumber \\
&+&\frac{1}{2}J_{AF_{1}}\mathbf{S}_{3j}\cdot \mathcal{S}_{3j+1}),
\label{LimitH}
\end{eqnarray}
where $\mathcal{S}_{j}$ is the spin operator with $S=1$. This model appears
to be rarely studied.

For arbitrary couplings, we will apply the DMRG \cite{DMRG1,DMRG2} method to
explore the ground-state properties of the tetrameric chain defined in Eq. (%
\ref{model Hamiltonian}). In the following DMRG calculations, the chain
length is taken as $N=160$, and the Hilbert space is truncated to $100$ most
relevant states. In the calculations for correlation function and string
order, $160$ optimal states are kept for accuracy. Open boundary conditions
are adopted. The truncation error is less than $10^{-8}$ in all calculations.

\section{Magnetization, Haldane-like gap, Correlation Function and String
order}

\subsection{Magnetization}

Due to the competition between interactions and the external magnetic field,
a quantum magnet often shows exotic properties under magnetic fields. A
field can close the zero-field gap, and may induce magnetization plateaux
under some conditions \cite{Hida,Okamoto,OYA,Gu2}. Therefore, we shall
investigate the magnetic properties of the present alternating tetrameric
system in a longitudinal field. Many cases with different couplings have
been calculated. Here the results of the case with $%
J_{AF_{1}}:J_{AF_{2}}:J_{F}=1:1:1$ are presented for example. Fig. \ref%
{magnetization} (a) shows the magnetic field dependence of the magnetization
per site $m(h)$. The system exhibits two magnetization plateaux, a
zero-field gap and a $m=1/4$ plateau, both of which satisfy the
Oshikawa-Yamanaka-Affleck (OYA) \cite{OYA} condition $n(S-m)=integer$, with $%
n$ the period of the ground state, $S$ the magnitude of spin and $m$ the
magnetization per site. A similar magnetization curve has been found in the $%
p$-merized antiferromagnetic spin chain with $p=4$ \cite{Cabra}. Both
systems have shown all plateaux that the OYA condition permits. The $m=0$
plateau in the present system exhibits a spin gap in the spectrum, which is
from the singlet ground state to the triplet excited states. Such a gap has
also been found in the spin-$1/2$ F-AF alternating Heisenberg chain \cite%
{Hida2}. In such spin-$1/2$ chains with even lattice translation symmetry,
the spectrum could have a gap even if the ground state has no spontaneous
translation symmetry breaking \cite{OYA,AL}. In contrast, spin-$1/2$ chains
with odd lattice sites per unit cell are gapless if the translation symmetry
does not break spontaneously, like the spin-$1/2$ Heisenberg chain \cite{LSM}
and the spin-$1/2$ F-F-AF trimerized chain \cite{Gu}. The further
discussions on this gapped ground states will be presented below.

\begin{figure}[tbp]
\includegraphics[width=1.0\linewidth,clip]{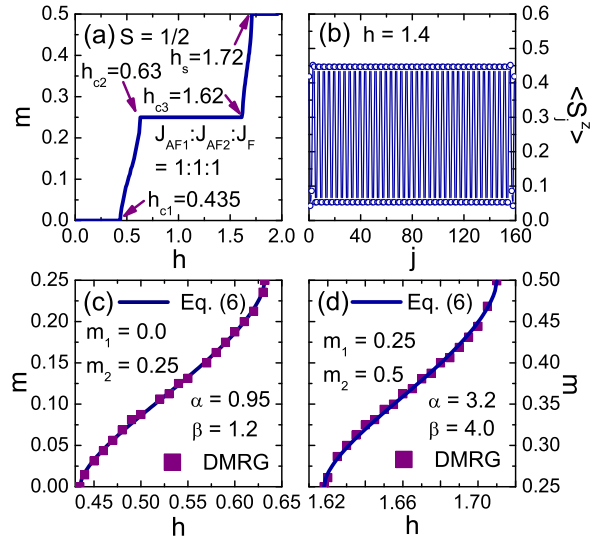}
\caption{(Color online) (a) The external field dependence of the
magnetization per site $m(h)$ for the spin-$1/2$ tetrameric HAFC with
couplings $J_{AF_{1}}$:$J_{AF_{2}}$:$J_{F}$=$1$:$1$:$1$. (b) The local
magnetic moment $\langle S^{z}_{j}\rangle$ in the magnetization plateau
states. The DMRG results of $m(h)$ in the gapless states can be fairly
fitted by Eq. (\protect\ref{fit}) for (c) $h_{c_{1}}<h<h_{c_{2}}$, and (d) $%
h_{c_{3}}<h<h_{s}$.}
\label{magnetization}
\end{figure}

For the $m=1/4$ plateau, the local magnetic moment $\langle S^{z}_{j}\rangle$
and the spin-spin correlation function $\langle S^{z}_{0}S^{z}_{j}\rangle$
are calculated. These two quantities show a perfect sequence with a period
of $4$ and $\langle S^{z}_{0}S^{z}_{j}\rangle$ has a long-range order. It is
shown that $\langle S^{z}_{j}\rangle$ behaves as $\left\lbrace\cdots,
(S_{1},S_{2},S_{2},S_{1}), \cdots \right\rbrace $ with $S_{1}=0.4468$ and $%
S_{2}=0.0532$, giving rise to the magnetization per site $m=1/4$, as shown
in Fig. \ref{magnetization} (b). The plateau state can be described by an
approximate wave function, 
\begin{eqnarray}
\psi_{i}&=&a\vert\uparrow_{4i-3}\uparrow_{4i-2}\uparrow_{4i-1}%
\downarrow_{4i} \rangle
+b\vert\uparrow_{4i-3}\uparrow_{4i-2}\downarrow_{4i-1}\uparrow_{4i} \rangle 
\nonumber \\
&+&c\vert\uparrow_{4i-3}\downarrow_{4i-2}\uparrow_{4i-1}\uparrow_{4i}
\rangle + d\vert\downarrow_{4i-3}\uparrow_{4i-2}\uparrow_{4i-1}\uparrow_{4i}
\rangle  \nonumber \\
(i&=&1,\cdots, N/4),
\end{eqnarray}
where $\uparrow_{j}$ ($\downarrow_{j}$) denotes spin up (down) on site $j$
and $N$ is the total site number. If the coefficients $a^{2}=d^{2}=S_{2}$
and $b^{2}=c^{2}=S_{1}$, the local magnetic moment and the spin-spin
correlation function deduced from this wave function perfectly fit into the
numerical results. The formation of the $m=1/4$ plateau in the limit $%
J_{AF_{1}}=0$ is quite simple, which is helpful for understanding the
general case. When $J_{AF_{1}}=0$, the system is decoupled into the isolated
dimers coupled by $J_{AF_{2}}$ or $J_{F}$. After applying a magnetic field,
the dimers coupled by $J_{F}$, which are in the triplet states, become
polarized and, the system exhibits a magnetization plateau at $m=1/4$. The
width of the plateau is determined by the upper critical field $J_{AF_{2}}$
leading to full polarization.

The magnetization process at $h_{c_{1}}<h<h_{c_{2}}$ and $h_{c_{3}}<h<h_{s}$
are shown in Figs. \ref{magnetization} (c) and (d), respectively. It is
found that the behaviors of the curves could be described as 
\begin{eqnarray}
m(h)&=&m_{1}+(h-h_{1})(m_{2}-m_{1})/(h_{2}-h_{1})  \nonumber \\
&+&\alpha(h_{2}-h)\sqrt{h-h_{1}}-\beta(h-h_{1})\sqrt{h_{2}-h},  \label{fit}
\end{eqnarray}
where $h_{1}$ ($h_{2}$) and $m_{1}$ ($m_{2}$) are the lower (upper) critical
field and the corresponding magnetization per site, $\alpha$ and $\beta$ are
two parameters. For $h_{c_{1}}<h<h_{c_{2}}$, $\alpha=0.95$ and $\beta=1.2$,
while for $h_{c_{3}}<h<h_{s}$, $\alpha=3.2$ and $\beta=4.0$. The curves
obtained from Eq. (\ref{fit}) are also shown in the figures, which fit into
the DMRG results quite well.

\subsection{Haldane-like gap}

To understand the properties of the ground state of the spin-1/2 tetrameric
alternating HAFC in the absence of a magnetic field, the spin gap, spin-spin
correlation function, and string order are studied by means of the DMRG
method in the whole parameter region, which will be compared with the
features of Haldane phase.

\begin{figure}[tbp]
\includegraphics[width=1.0\linewidth,clip]{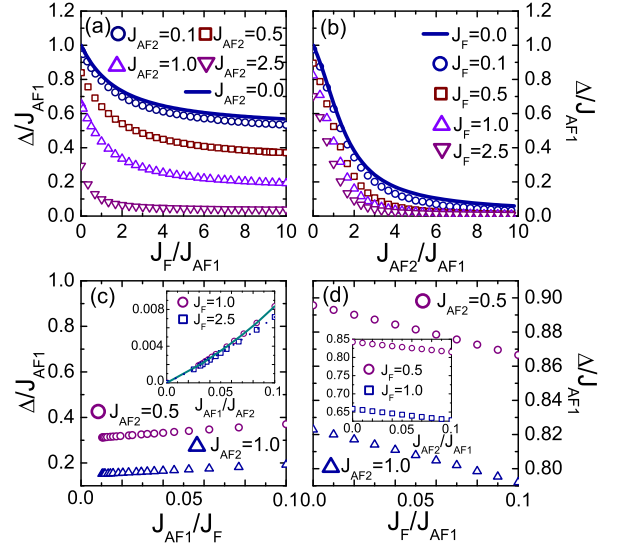}
\caption{(Color online) (a) The $J_{F}$ dependence of the gap $\Delta$ for $%
J_{AF_{2}}$=$0.0,0.1,0.5,1.0$ and $2.5$. (b) The $J_{AF_{2}}$ dependence of
the gap $\Delta$ for $J_{F}$=$0.0,0.1,0.5,1.0$ and $2.5$. (c) The gap
behaviors with $J_{AF_{1}}$/$J_{F}$ for large $J_{F}$. The inset shows the
gap behaviors with $J_{AF_{1}}$/$J_{AF_{2}}$ for large $J_{AF_{2}}$. (d) The
gap behaviors with $J_{F}$ for large $J_{AF_{1}}$. The inset shows the gap
behaviors with $J_{AF_{2}}$ for large $J_{AF_{1}}$.}
\label{GAP}
\end{figure}

We consider the spin gap $\Delta$ from the ground state to the triplet
excited states, namely, 
\begin{equation}
\Delta=E_{1}-E_{0},
\end{equation}
where $E_{0}$ is the ground-state energy and $E_{1}$ is the lowest energy in
the subspace with $S^{z}_{tot}=1$, and $J_{AF_{1}}$ is chosen as the energy
scale. $J_{F}$ and $J_{AF_{2}}$ dependences of the gap are presented in
Figs. \ref{GAP} (a) and (b), respectively. Fig. \ref{GAP} (a) shows the gap
as a function of $J_{F}$ for $J_{AF_{2}}= 0.0,0.1,0.5,1.0$ and $2.5$. When $%
J_{AF_{2}}=0$, the gap is determined by Eq. (\ref{gap_Jf}). It is shown that
with the increase of $J_{F}$, the gap $\Delta$ smoothly decreases and
converges to a certain value $\Delta_{c}$ when $J_{F}\rightarrow \infty$. $%
\Delta_{c}$ is determined by $J_{AF_{2}}$. In the limit of $J_{AF_{2}}=0$, $%
\Delta_{c}=J_{AF_{1}}/2$. In the large $J_{F}$ limit, the system is
equivalent to the spin-($1$,$1/2$,$1/2$) model, which will be proved to have
a gap by the non-linear $\sigma$ model below, where $\Delta_{c}$ is related
to the gap of this model. With the increase of $J_{F}$, the tetrameric
system changes from a tetramer model to the spin-($1$,$1/2$,$1/2$) model,
and the gap decreases continuously. In Fig. \ref{GAP} (b), the gap is
plotted as a function of $J_{AF_{2}}$ for $J_{F}=0.0,0.1,0.5,1.0$ and $2.5$.
When $J_{F}=0$, the gap is evaluated by Eq. (\ref{gap_J2}). It is observed
that the gap decreases rapidly with increasing $J_{AF_{2}}$ and tends to
zero when $J_{AF_{2}}\rightarrow \infty$. With the increase of $J_{AF_{2}}$,
the tetrameric chain changes from a tetramer model to a system with relative
small $J_{AF_{1}}$ interactions, which is gapless when $J_{AF_{1}}=0$. From
the above discussions, it can be seen that both $J_{F}$ and $J_{AF_{2}}$
give rise to the decrease of the gap, but with different behaviors. The
influence of $J_{AF_{1}}$ on the gap is found to be distinct from those of $%
J_{AF_{2}}$ and $J_{F}$. The gap would increase with $J_{AF_{1}}$. Because
the gap only vanishes when $J_{AF_{2}}/J_{AF_{1}}\rightarrow \infty$, it is
reasonable to conclude that a gap would be generated by an arbitrary small $%
J_{AF_{1}}$.

Next, let us consider the asymptotic behaviors of the gap. In Fig. \ref{GAP}
(c), the gap as a function of $J_{AF_{1}}/J_{F}$ is plotted in large $J_{F}$
limit for $J_{AF_{2}}=0.5$ and $1.0$. It is shown that the gaps linearly
converge to $\Delta _{c}$. $J_{AF_{2}}$ alters the value of $\Delta _{c}$
but does not change this linear asymptotic behavior. Thus, the variations of
the gap for large $J_{F}$ could be evaluated by 
\begin{equation}
\Delta =J_{AF_{1}}(\frac{\Delta _{c}}{J_{AF_{1}}}+\mu \frac{J_{AF_{1}}}{J_{F}%
}),
\end{equation}%
where $\Delta _{c}$ and the parameter $\mu $ are determined by $J_{AF_{2}}$.
In the large $J_{AF_{2}}$ limit, the isolated tetramer system with $J_{F}=0$
gives $\Delta \simeq J_{AF_{1}}^{2}/J_{AF_{2}}$. The inset of Fig. \ref{GAP}
(c) shows the gap behaviors for $J_{F}=1.0$ and $2.5$ for large $J_{AF_{2}}$%
, which exhibits an asymptotic behavior 
\begin{equation}
\Delta \sim \frac{J_{AF_{1}}^{2}}{J_{AF_{2}}}e^{\gamma
J_{AF_{1}}/J_{AF_{2}}},
\end{equation}%
where $\gamma $ is a parameter, and $\gamma =3.2$ for $J_{F}=1.0$ and $1.8$
for $J_{F}=2.5$. The extrapolations show the linear behaviors near $%
J_{AF_{1}}=0$. When $J_{AF_{1}}$ dominates, the gap behaviors are shown in
Fig. \ref{GAP} (d). It can be seen that the gap decreases linearly with
increasing both $J_{AF_{2}}$ and $J_{F}$ for large $J_{AF_{1}}$.

The DMRG results suggest that the system is in disordered spin liquid states
with a spin gap. The gap varies monotonically with the couplings,
and does not show any singularity. We have calculated the ground state
energy against the couplings and found that the ground state energy varies
also monotonically with the interactions, and there is no any
nonanalyticity. This nonanalyticity of the ground state energy suggests an
absence of the quantum phase transition in this system \cite{QFT}.

It has been shown that the gap always exists if $J_{AF_{1}}$ is finite.
However, it is hard to identify the validity of this statement from
numerical calculations when $J_{AF_{2}}\gg J_{AF_{1}}$. Thus, we consider
the nonlinear $\sigma $ model (NLSM) of this tetrameric system. Because the
gap decreases with $J_{F}$, we consider the case in the limit of $%
J_{F}\rightarrow \infty $, when this tetrameric system is equivalent to the
spin-($1$,$1/2$,$1/2$) model which is described by Eq. (\ref{LimitH}). The
NLSM for such spin chains has been developed by Affleck \cite{NLSM1}, Fukui
and Kawakami \cite{NLSM2}, and Takano \cite{NLSM3}. The magnitude of $%
\mathbf{S}_{j}$ is denoted as $s_{j}$, which could be $1/2$ or $1$ here. The
period of the system $2b$ is regarded as $6$ for convenience. After a
standard procedure \cite{NLSM3}, we have the effective action for the spin-($%
1$,$1/2$,$1/2$) Hamiltonian: 
\begin{eqnarray}
S_{eff} &=&\int_{0}^{\beta }d\tau \int_{0}^{L}dx\{\frac{1}{2aJ^{(1)}}(\frac{%
J^{(1)}}{J^{(2)}}-\frac{J^{(0)}}{J^{(1)}})(\partial _{\tau }\mathbf{m})^{2} 
\nonumber \\
&-&i\frac{J^{(0)}}{J^{(1)}}\mathbf{m}\cdot \left( \partial _{\tau }\mathbf{m}%
\times \partial _{x}\mathbf{m}\right) +\frac{aJ^{(0)}}{2}(\partial _{x}%
\mathbf{m})^{2}\},
\end{eqnarray}%
where $1/\beta $ denotes the temperature, $a$ is the lattice spacing, $L$ is
the length of the chain, $\mathbf{m}$ presents the spin variables, and $%
\left\{ J^{n}\right\} $ are defined by 
\begin{equation}
\frac{1}{J^{(n)}}=\frac{1}{2b}\sum_{q=1}^{2b}\frac{(\tilde{s}_{q})^{n}}{%
\tilde{J}_{q}},\quad (n=0,1,2)
\end{equation}%
with $\tilde{J}_{q}=J_{q}s_{q}s_{q+1}$ and $\tilde{s}_{q}=%
\sum_{k=1}^{q}(-1)^{k+1}s_{k}$. The topological angle $\theta $ is $4\pi
J^{(0)}/J^{(1)}$ and the NLSM provides a gapless condition when $\theta
/2\pi $ is a half-odd integer. After applying these equations to the spin-($%
1 $,$1/2$,$1/2$) model, we have the gapless condition 
\begin{equation}
\frac{J_{AF_{1}}}{J_{AF_{2}}}=\frac{6l-11}{6-4l}\quad (l=positive\ integer).
\label{gapless}
\end{equation}%
As $J_{AF_{1}}$ and $J_{AF_{2}}$ are both positive, Eq. (\ref{gapless})
could not be satisfied in any case, which indicates that even in the large $%
J_{F}$ limit, the present tetrameric chain is gapped for any $J_{AF_{1}}>0$
and $J_{AF_{2}}>0$. Considering the effect of $J_{F}$ on the gap, we could
conclude that this spin-1/2 tetrameric Heisenberg AF chain is always gapped
if $J_{AF_{1}}\neq 0$.

\begin{figure}[tbp]
\includegraphics[width=1.0\linewidth,clip]{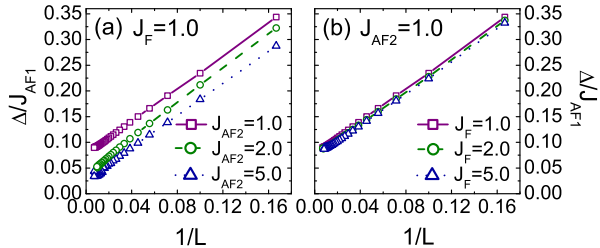}
\caption{(Color online) The gap $\Delta$ of the spin-$5/2$ tetrameric HAFC
as a function of the inverse of chain length $(1/L)$ for (a) $J_{F}=1.0$, $%
J_{AF_{2}}$=$1.0,2.0$ and $5.0$; and (b) $J_{AF_{2}}=1.0$, $J_{F}=1.0,2.0$
and $5.0$.}
\label{5/2}
\end{figure}

For a comparison, the gap of the spin-$5/2$ tetrameric HAFC is
computed by utilizing the DMRG method. The chain length is taken as $140$
and the Hilbert space is truncated to $400$ optimal states. The truncation
error is less than $10^{-13}$ in energy calculations. In Fig. \ref{5/2}, the
DMRG results show that the extrapolations of the gap converge to nonzero
values. Similar to the case with spin-$1/2$, the gap of this spin-$5/2$
tetrameric HAFC also decreases with both increasing $J_{AF_{2}}$ and $J_{F}$%
. The gap diminishes more rapidly with increasing $J_{AF_{2}}$, which is
also a feature of the spin-$1/2$ case. Although the gap of the spin-$5/2$
system is smaller than that in the spin-$1/2$ case, the qualitative
behaviors of the gap for both systems appear to be similar.

\subsection{Correlation function and static structure factor}

To characterize the gapped ground states of the tetrameric spin chain, let
us investigate the spin-spin correlation function $\langle
S^{z}_{i}S^{z}_{j}\rangle$ and the static structure factor $S(q)$ that is
defined as the Fourier transform of the correlation function: 
\begin{equation}
S(q)=\frac{1}{N}\sum_{i,j}\langle S^{z}_{i}S^{z}_{j}\rangle e^{iq(i-j)}.
\end{equation}
Our DMRG results show that the correlation function decays exponentially in
the gapped states. For different couplings, the correlation length and the
behavior of the correlation function have dramatic changes, which could be
displayed more clearly in the static structure factor $S(q)$.

\begin{figure}[tbp]
\includegraphics[width=1.0\linewidth,clip]{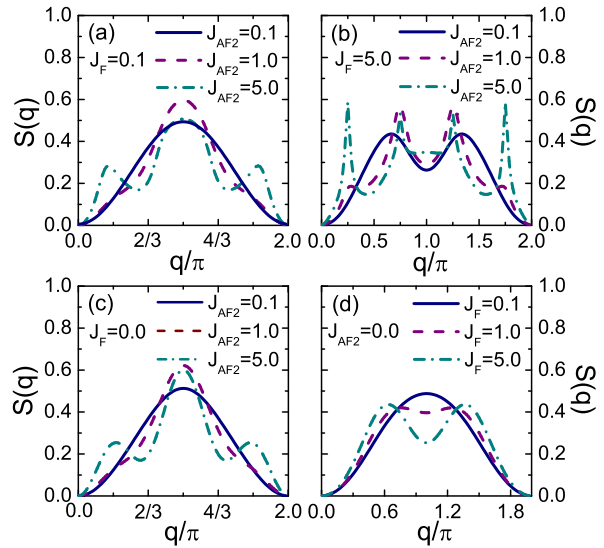}
\caption{(Color online) The DMRG results of the static structure factor $%
S(q) $ in the gapped ground states of the spin-$1/2$ tetrameric HAFC for (a) 
$J_{F}$=$0.1$, $J_{AF_{2}}$=$0.1,1.0$ and $5.0$; (b) $J_{F}$=$5.0$, $%
J_{AF_{2}}$=$0.1,1.0$ and $5.0$; (c) $J_{F}$=$0.0$, $J_{AF_{2}}$=$0.1,1.0$
and $5.0$; (d) $J_{AF_{2}}$=$0.0$, $J_{F}$=$0.1,1.0$ and $5.0$.}
\label{Sq}
\end{figure}

We have performed a large amount of calculations on the correlation function
and static structure factor in a wide range of the parameter region. The
changes of $S(q)$ are shown in Figs. \ref{Sq} (a) and (b) for some
parameters as examples. In Fig. \ref{Sq} (a), for $J_{F}=0.1$, $%
J_{AF_{2}}=0.1,1.0$ and $5.0$, $S(q)$ shows an obvious maximum at $q=\pi$,
and with the enhancement of $J_{AF_{2}}$, there are two small maxima
appearing at $q=\pi/3$ and $5\pi/3$. When $J_{AF_{2}}=0.1$ and $J_{F}=5.0$,
the maximum at $q=\pi$ becomes a valley and two small maxima appear near $%
q=3\pi/5$ and $7\pi/5$. With further increasing $J_{AF_{2}}$, $S(q)$
exhibits a totally different behavior, showing four peaks at $%
q=\pi/4,3\pi/4,5\pi/4$ and $7\pi/4 $, as shown in Fig. \ref{Sq} (b). In
these numerical results, $S(q)$ shows three different behaviors due to the
competition of the couplings. It will be shown that some features of $S(q)$
mentioned above are determined by the short-range correlations, and some
others are due to the increase of the correlation length and the translation
symmetry of the ground states.

The correlation function of this model has no long-range order, but exhibits
a short-range order characterized by some $q_{max}$ of $S(q)$. In the limit $%
J_{AF_{2}}=0$ and $J_{F}=0$, the tetrameric chain is decoupled into singlet
dimers, and the static structure factor has a simple form $S(q)=(1-\cos q)/4$%
. The short-range correlations $\langle S_{2i-1}^{z}S_{2i}^{z}\rangle $ are
reflected by the maximum of $S(q)$ at $q=\pi $. When $J_{AF_{2}}$ is
switched on, the tetrameric chain becomes a tetramer system. The short range
correlations of the tetramer produce another two maxima of $S(q)$ at $q=\pi
/3$ and $5\pi /3$, as shown in Fig. \ref{Sq} (c). On the other hand, when $%
J_{F}$ is turned on, the short-range correlations drive the maximum of $S(q)$
at $q=\pi $ to split into two small maxima, which move towards $q$=$3\pi /5$
and $7\pi /5$ with increasing $J_{F}$, as presented in Fig. \ref{Sq} (d).
The distinct behaviors of $S(q)$ in the two situations are due to the
different couplings of the tetramers. These results found in the tetramer
systems also explain the behaviors of $S(q)$ for small $J_{AF_{2}}$ or $%
J_{F} $ shown in Figs. \ref{Sq} (a) and (b). When $J_{AF_{2}}$ or $J_{F}$ is
smaller than $J_{AF_{1}}$, the correlation length of the system is quite
short, reflecting a highly disordered ground state, and the correlation
function and static structure factor could be well characterized by the
corresponding decoupled tetramer system.

When both $J_{AF_{2}}$ and $J_{F}$ are large enough, $S(q)$ exhibits a
totally different behavior. There are four peaks at $q=\pi /4,3\pi /4,5\pi /4
$ and $7\pi /4$, as shown in Fig. \ref{Sq} (b) when $J_{AF_{2}}=J_{F}=5.0$.
In this case, the correlation function keeps exponentially decaying but
shows a structure with a period of $4$, and its correlation length becomes
large. It is also observed that the correlation function has a
characteristic such as $\langle S_{i}^{z}S_{j}^{z}\rangle \simeq \langle
S_{i}^{z}S_{j+1}^{z}\rangle $ if the spins $\mathbf{S}_{j}$ and $\mathbf{S}%
_{j+1}$ are coupled by $J_{F}$, and $\langle S_{i}^{z}S_{j}^{z}\rangle
\simeq -\langle S_{i}^{z}S_{j+1}^{z}\rangle $ if the spins $\mathbf{S}_{j}$
and $\mathbf{S}_{j+1}$ are coupled by $J_{AF_{2}}$. This feature reflects
that the spins coupled by $J_{F}$ behave like triplets, and the spins
coupled by $J_{AF_{2}}$ behave like singlets. As the correlation
function that decays exponentially is modulated with a period of $4$, the
static structure factor $S(q)$ shows the peaks at $q=\pi /4,3\pi /4,5\pi /4$
and $7\pi /4$. Such a modulated structure of $S(q)$ has also been observed in the spin-%
$1/2$ alternating F-AF \cite{FAF} and spin-$1/2$ trimerized F-F-AF
Heisenberg chains \cite{GSS}, which show the peaks of $S(q)$ at $q=\pi
/2,3\pi /2$ and $q=\pi /3,\pi ,5\pi /3$, respectively. Generally speaking,
for an exponentially decaying correlation function that is modulated with a
period of $n$, the static structure factor $S(q)$ could show
antiferromagnetic peaks at $q=\pi (1+2l)/n$ with $l=0,1,2,\cdots ,n-1$.

Although the ground state energy has no nonanalyticity in the
parameter space, namely, no quantum phase transition happens, the static
structure factor $S(q)$ shows rich features of the ground states for
different couplings in this tetrameric system.

\subsection{Topological string order}

In order to investigate natures of the Haldane-like phase in the disordered
phase of the tetrameric chain, we have calculated the string order to detect
the hidden symmetry in the gapped phase. The nonlocal string order was found
in spin-$1$ antiferromagnetic Heisenberg chain by Den Nijs and Rommelse \cite%
{NR}. They found that although the spins with $S^{z}_{i}=1,0,-1$ are not
ordered in position, their sequence has a hidden N\'{e}el order, that is, if
all sites with $S^{z}_{i}=0$ are removed, the left sites with $%
S^{z}_{i}=1,-1 $ have a N\'{e}el order. The string order was found to be a
common feature of Haldane phase. Hida suggested a string order for spin-$1/2$
F-AF chain \cite{Hida2}, which was used to characterize the phase diagram of
this spin chain \cite{YHK}. The string order for spin ladders has also been
considered in distinct gapped phases \cite{SV}.

Here we suggest a string order for this tetrameric system, and use it to
characterize the disordered spin liquid phase. Considering the basic feature
of string order and the translation symmetry of the tetrameric Hamiltonian,
we define four string operators for this tetrameric chain, which are denoted
by $\Theta^{\alpha}_{4i-3,4j+2}$, $\Theta^{\alpha}_{4i-2,4j+3}$, $%
\Theta^{\alpha}_{4i-1,4j}$, and $\Theta^{\alpha}_{4i,4j+1}$. $%
\Theta^{\alpha}_{4i-3,4j+2}$ is defined as 
\begin{eqnarray}
\Theta^{\alpha}_{4i-3,4j+2}=-S^{\alpha}_{4i-3}\exp(i\pi%
\sum_{k=4i-2}^{4j+1}S^{\alpha}_{k})S^{\alpha}_{4j+2},
\end{eqnarray}
and the corresponding order parameter is 
\begin{equation}
O^{\alpha}_{4i-3,4j+2}=\lim_{|j-i|\rightarrow\infty}\left\langle
\Theta^{\alpha}_{4i-3,4j+2}\right\rangle ,
\end{equation}
where $i$ and $j$ denote the unit cell and $\alpha=x,y$ or $z$. The other
three ones are defined in the similar way and they could be obtained by a
translation of $\Theta^{\alpha}_{4i-3,4j+2}$. As the present model in the
absence of a magnetic field has a $SU(2)$ symmetry, we need only to consider 
$\Theta^{z}$. The spin configuration shows that $\Theta^{z}_{4i-3,4j+2}$ and 
$\Theta^{z}_{4i-1,4j}$ have the same topological structure, while $%
\Theta^{z}_{4i-2,4j+3}$ and $\Theta^{z}_{4i,4j+1}$ possess another same
structure. The DMRG calculations have been performed on the four string
orders. It was found that the string orders of the first structure have the
same finite value, while the string orders of the second structure are both
zero in the whole parameter space. For the spin-$1/2$ F-AF alternating
chain, the two different definitions of string order also have two different
values, one finite and one zero \cite{SNT}. This difference would be
explained in the next section. Here we only consider the nonvanishing string
order, which would be denoted as $O^{z}_{string}$ below.

\begin{figure}[tbp]
\includegraphics[width=1.0\linewidth]{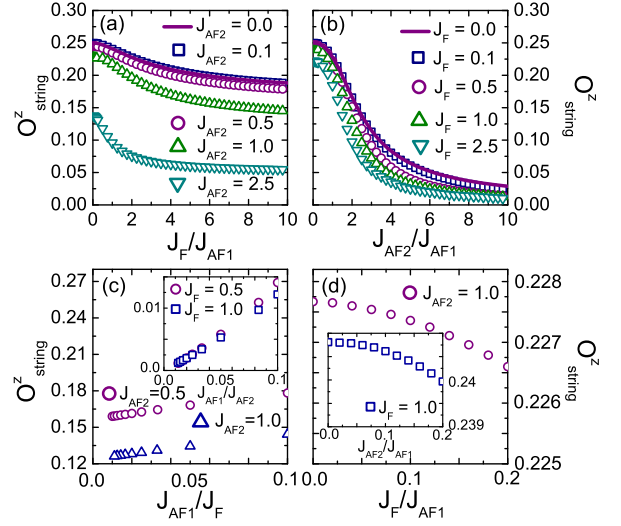}
\caption{(Color online) The string order as a function of $J_{F}/J_{AF_{1}}$
(a) and $J_{AF_{2}}/J_{AF_{1}}$ (b). The asymptotic behavior of the string
order for large $J_{F}$ or $J_{AF_{2}}$ (c) and for large $J_{AF_{1}}$ (d).}
\label{String}
\end{figure}

When both $J_{AF_{2}}$ and $J_{F}$ are equal to zero, the string order has
its maximum $0.25$, as shown in Fig. \ref{String}. After tuning up $J_{F}$
or $J_{AF_{2}}$, the system becomes tetramer assemblies. In the limit $%
J_{AF_{2}}=0$, the string order is evaluated from the wavefunction of the
tetramer ground state as a function of $J_{F}/J_{AF_{1}}$: 
\begin{equation}
O^{z}_{string}=\frac{4+\frac{J_{F}}{J_{AF_{1}}}+\sqrt{4+2\frac{J_{F}}{%
J_{AF_{1}}}+(\frac{J_{F}}{J_{AF_{1}}})^{2}}}{12\sqrt{4+2\frac{J_{F}}{%
J_{AF_{1}}}+(\frac{J_{F}}{J_{AF_{1}}})^{2}}},
\end{equation}
while for $J_{F}=0$, the string order as a function of $%
J_{AF_{2}}/J_{AF_{1}} $ is expressed as 
\begin{equation}
O^{z}_{string}=\frac{4-\frac{J_{AF_{2}}}{J_{AF_{1}}}+\sqrt{4-2\frac{%
J_{AF_{2}}}{J_{AF_{1}}}+(\frac{J_{AF_{2}}}{J_{AF_{1}}})^{2}}}{12\sqrt{4-2%
\frac{J_{AF_{2}}}{J_{AF_{1}}}+(\frac{J_{AF_{2}}}{J_{AF_{1}}})^{2}}},
\end{equation}
as shown with solid lines in Figs. \ref{String} (a) and (b), respectively.
From these equations and numerical simulations, one may find that the string
order in the case of $J_{AF_{2}}=0$ has an asymptotic behavior:
\begin{equation}
O^{z}_{string} \sim \frac{1}{4}-\frac{1}{64}(\frac{J_{F}}{J_{AF_{1}}}%
)^{2}e^{-J_{F}/2J_{AF_{1}}},  \label{c1}
\end{equation}
when $J_{F}\ll J_{AF_{1}}$. In the case of $J_{F}=0$, it behaves as 
\begin{equation}
O^{z}_{string} \sim \frac{1}{4}-\frac{1}{64}(\frac{J_{AF_{2}}}{J_{AF_{1}}}%
)^{2}e^{J_{AF_{2}}/2J_{AF_{1}}},  \label{c2}
\end{equation}
when $J_{AF_{2}}\ll J_{AF_{1}}$. On the other hand, in the limit of $%
J_{AF_{2}}=0$, the string order has an asymptotic form when $J_{F}\gg
J_{AF_{1}}$, 
\begin{equation}
O^{z}_{string} \sim \frac{1}{6}+\frac{1}{4}\frac{J_{AF_{1}}}{J_{F}}%
e^{-2J_{AF_{1}}/J_{F}},  \label{c3}
\end{equation}
while it behaves as 
\begin{equation}
O^{z}_{string} \sim \frac{1}{4}\frac{J_{AF_{1}}}{J_{AF_{2}}}%
e^{2J_{AF_{1}}/J_{AF_{2}}},  \label{c4}
\end{equation}
when $J_{AF_{2}}\gg J_{AF_{1}}$ for the case $J_{F}=0$.

Besides the limiting cases, the string order is evaluated as a function of
the couplings by means of the DMRG method to detect the hidden symmetry in
the gapped phase. Fig. \ref{String} presents the DMRG results of the string
order as functions of $J_{F}/J_{AF_{1}}$ and $J_{AF_{2}}/J_{AF_{1}}$. In
Fig. \ref{String} (a), the string order is shown as a function of $%
J_{F}/J_{AF_{1}}$ for $J_{AF_{2}}=0.1,0.5,1.0$ and $2.5$. The string order
decreases monotonically with $J_{F}/J_{AF_{1}}$ and converges to a nonzero
value when $J_{F}/J_{AF_{1}}\rightarrow \infty$. In Fig. \ref{String} (b),
the string order as a function of $J_{AF_{2}}/J_{AF_{1}}$ is evaluated for $%
J_{F}=0.1,0.5,1.0$ and $2.5$. It is shown that it decreases rapidly with $%
J_{AF_{2}}/J_{AF_{1}}$ and vanishes when $J_{AF_{2}}/J_{AF_{1}}\rightarrow
\infty$. As is seen, the string order is always finite in the gapped phase,
indicating the existence of a hidden symmetry in this phase. The variations
of the string order show the same tendencies as the gap behaviors. The
string order has its maximum when $J_{F}=0$ and $J_{AF_{2}}=0$, where the
system also shows the maximum of the gap. When the gap vanishes in the case
of $J_{AF_{1}}=0$, the string order is also zero. This indicates that the
opening of the gap and its behaviors could be characterized by this string
order.

The asymptotic behaviors of the string order are shown in Figs. \ref{String}
(c) and (d). Fig. \ref{String} (c) shows the string orders when $J_{F}$$\gg$$%
J_{AF_{1}}$ for $J_{AF_{2}} =0.5$ and $1.0$. The inset shows the string
orders when $J_{AF_{2}}$$\gg$$J_{AF_{1}}$ for $J_{F}=0.5$ and $1.0$. Fig. %
\ref{String} (d) and the inset show the string orders when $J_{AF_{1}}$$\gg$$%
J_{F}$ and $J_{AF_{1}}$$\gg$$J_{AF_{2}}$, respectively. It is found that the
asymptotic behaviors of the string orders preserve the features as those of
the corresponding tetramer models, which are described by Eqs. (\ref{c1})-(%
\ref{c4}).

It should be noticed that although the string order is nonlocal, it could
measure some localized singlet correlation that depicts the singlet state.
The string order has been found to characterize the behaviors of the gap
perfectly, as shown in Figs. \ref{String} (a) and (b). The gap monotonically
increases with $J_{AF_{1}}$, and is related to the singlet states of the
spins coupled by $J_{AF_{1}}$, which could also be measured by the string
order. When the system is the assembly of localized singlet spins coupled by 
$J_{AF_{1}}$, the string order has its maximum value $0.25$. With the
disappearance of the singlet spins for $J_{AF_{1}}=0$, the string order also
vanishes. In the spin-$1/2$ F-AF alternating chain, the gap is also related
to the singlet dimers coupled by the AF interactions, and the string order
could measure the singlet correlation \cite{Hida2}. Hida \cite{Hida2}
pointed out that such a string order could distinguish the valence-bond-type
disordered states from other disordered states. In this tetrameric chain,
the valence-bonds, namely the singlet states in the spins coupled by $%
J_{AF_{1}}$, is characterized by the string order.

The finite string order reveals the hidden N\'{e}el order in the gapped
states of the tetrameric chain, which is regarded as a feature of Haldane
phase. For this tetrameric model, the valence-bond state could provide us a
picture for the hidden N\'{e}el order. According to our definition of the
string order, the exponent part of the string operator, which is between the
spins at two boundaries, could be regarded as those constructed by the spin
pairs coupled by $J_{AF_{2}}$ or $J_{F}$. Consequently, the valence bonds
form between spins coupled by $J_{AF_{1}}$, leading to that the spin pairs
coupled by $J_{AF_{2}}$ or $J_{F}$ could have $S_{tot}=1,0,-1$. After
removing the spin pairs with $S_{tot}=0$, we can see a perfect N\'{e}el
order of the spin pairs.

\section{Duality and hidden symmetry}

In the Haldane phase of the spin-$1$ Heisenberg chain \cite{KT} and spin-$%
1/2 $ F-AF alternating chain \cite{MH}, the hidden symmetry measured by
string order was found to be a $Z_{2}$ symmetry after a dual transformation,
and the string order is transformed into a ferromagnetic order to measure
such a $Z_{2}$ symmetry in the dual space. In general, the string orders in
both $x$ and $z$ axes are considered to construct a $Z_{2}\times Z_{2}$
symmetry of the transformed Hamiltonian. The hidden symmetry could be
revealed by means of the dual transformation \cite{KNK,MH} and the Haldane
phase was characterized by a fully breaking of the $Z_{2}\times Z_{2}$
symmetry \cite{KT,YHK}. It is interesting that this tetrameric chain also
shows a hidden $Z_{2}\times Z_{2}$ symmetry, which is just measured by the
string order defined above and fully breaking in this gapped phase. For
simplicity, we start with the Hamiltonian rewritten in terms of Pauli
matrices 
\begin{eqnarray}
H &=&\sum\limits_{j}(J_{AF_{1}}\mathbf{\sigma}_{4j-3}\cdot \mathbf{\sigma}%
_{4j-2}+J_{AF_{2}}\mathbf{\sigma} _{4j-2}\cdot \mathbf{\sigma}_{4j-1} 
\nonumber \\
&+&J_{AF_{1}}\mathbf{\sigma}_{4j-1}\cdot \mathbf{\sigma}_{4j}-J_{F}\mathbf{%
\sigma}_{4j}\cdot \mathbf{\sigma}_{4j+1}).
\end{eqnarray}
After applying the standard Kramers-Wannier dual transformation $D$ \cite{KW}
to the whole system, we get the transformed operators 
\begin{eqnarray}
D\sigma^{x}_{j}\sigma^{x}_{j+1}D^{-1}&=&\sigma^{z}_{j-1+1/2}%
\sigma^{z}_{j+1+1/2},  \nonumber \\
D\sigma^{y}_{j}\sigma^{y}_{j+1}D^{-1}&=&-\sigma^{z}_{j-1+1/2}%
\sigma^{x}_{j+1/2}\sigma^{z}_{j+1+1/2},  \nonumber \\
D\sigma^{z}_{j}\sigma^{z}_{j+1}D^{-1}&=&\sigma^{x}_{j+1/2},
\end{eqnarray}
where $\sigma^{\alpha}_{j+1/2}$ ($\alpha =x,y,z$) are the Pauli operators in
the dual space. In order to present the spins more clearly, we merely change
the labeling of the lattice sites by the following rule: 
\begin{equation}
R: r\rightarrow \frac{1}{2}(r+2-\frac{1}{2}).
\end{equation}
After relabeling, we write the Pauli matrices on the half-odd-integer spins
as $\tau$ and apply the inverse of the Kramers-Wannier dual transformation
only on $\tau$ spins. As a result, the odd sites of both $\sigma$ and $\tau$
spins are rotated by $\pi$ about the $x$ axis. Thus, the transformed
Hamiltonian has the form of 
\begin{eqnarray}
\tilde{H}&=&-J_{AF_{1}}\sum_{j}(\sigma^{z}_{j}\sigma^{z}_{j+1}+\tau^{z}_{j}%
\tau^{z}_{j+1}+\sigma^{z}_{j}\sigma^{z}_{j+1}\tau^{z}_{j}\tau^{z}_{i+1}) 
\nonumber \\
&+&\sum_{j}(J_{AF_{2}}\sigma^{x}_{2j}+J_{F}\sigma^{x}_{2j+1})+%
\sum_{j}(J_{AF_{2}}\tau^{x}_{2j}+J_{F}\tau^{x}_{2j+1})  \nonumber \\
&+&\sum_{j}(J_{F}\sigma^{x}_{2j+1}\tau^{x}_{2j+1}-J_{AF_{2}}\sigma^{x}_{2j}%
\tau^{x}_{2j}).  \label{AT}
\end{eqnarray}
The tetrameric Heisenberg chain is transformed into a quantum Ashkin-Teller
(AT) model in a transverse field which is described by Eq. (\ref{AT}). This
AT model consists of two Ising chains coupled by four-component
interactions. The transverse field parts measured by $J_{AF_{2}}$ and $J_{F}$
in this model play the role of temperature in classical systems. They
compete with $J_{AF_{1}}$ to determine the behavior of the system, which has
been demonstrated in our numerical results. Similar to the quantum Ising
model \cite{QFT}, a phase transition, which is referred to as a spontaneous
breaking of the $Z_{2}$ symmetry and measured by the ferromagnetic order,
may happen because of these competitions. The symmetry of this quantum AT
model can be easily read off. It is invariant under rotations of $\pi$ about
the $x$ axis that are applied to $\sigma$ spins alone or $\tau$ spins alone.
Thus, it has a $Z_{2}\times Z_{2}$ symmetry.

After the same dual transformation, the difference of the four string
operators defined above is revealed. The vanishing string orders are
transformed into 
\begin{eqnarray}
U\Theta^{z}_{4i-2,4j+3}U^{-1}&=&-\otimes^{2j+2}_{k=2i}\sigma^{x}_{k}, 
\nonumber \\
U\Theta^{z}_{4i,4j+1}U^{-1}&=&-\otimes^{2j+1}_{k=2i+1}\sigma^{x}_{k},
\end{eqnarray}
where $U$ denotes the whole transformation. These two string operators are
still expressed in terms of nonlocal operators and could not measure the
spontaneous breaking of the discrete symmetry. In contrast, the nonvanishing
string orders become ferromagnetic correlations in the dual space, 
\begin{eqnarray}
U\Theta^{z}_{4i-3,4j+2}U^{-1}&=&(-1)^{j-i}\tau^{z}_{2i-1}\tau^{z}_{2j+2}, 
\nonumber \\
U\Theta^{z}_{4i-1,4j}U^{-1}&=&(-1)^{j-i+1}\tau^{z}_{2i}\tau^{z}_{2j+1},
\end{eqnarray}
where the factors $(-1)^{j-i}$ and $(-1)^{j-i+1}$ come from the rotation
transformation on the odd spins. These order parameters measure possible
spontaneous breaking of the $Z_{2}$ symmetry of $\tau$ spins. The
ferromagnetic order of $\sigma$ spins corresponds to the string order
defined in the $x$ component. When both $J_{AF_{2}}=0$ and $J_{F}=0$, the AT
model becomes two ferromagnetic Ising chains coupled by the four-component
interactions. The ferromagnetic correlation has the maximum $1$, and the $%
Z_{2}\times Z_{2}$ symmetry is totally broken because of the full
polarization along the $z$ axis. In contrast, when $J_{AF_{1}}=0$, the AT
model becomes isolated rungs in a transverse field. The ferromagnetic
correlation is $0$, and the $Z_{2}\times Z_{2}$ symmetry is preserved. In
our model, we are considering a $SU(2)$ symmetric model, thus the $x$
component of the string order equals to that of the $z$ component, namely $%
O^{x}=O^{z}$. It turns out that, $O^{z}$ could also be used to measure the $%
Z_{2}$ symmetry of $\sigma$ spins. The nonvanishing string order $O^{z}$
indicates that the $Z_{2}\times Z_{2}$ symmetry are fully breaking in the
tetrameric system, which is an important evidence for the Haldane-like type
of the gapped phase.

In this quantum AT model, the different effects of the couplings $J_{AF_{2}}$
and $J_{F}$ could be seen more clearly. As shown in the transformed AT
Hamiltonian Eq. (\ref{AT}), the odd-site spins have an antiferromagnetic
interaction $J_{F}\sigma^{x}_{2j+1}\tau^{x}_{2j+1}$, which has an inverse
effect on the transverse field part $J_{F}(\sigma^{x}_{2j+1}+%
\tau^{x}_{2j+1}) $, and thus counteracts the polarization of the odd-site
spins on the $x$ direction. In contrast, the even-site spins have a
ferromagnetic interaction $-J_{AF_{2}}\sigma^{x}_{2j}\tau^{x}_{2j}$, which
promotes the polarization on the $x$ direction. This difference could partly
explain why the gap and string order would vanish when $J_{AF_{2}}$$%
\rightarrow\infty$, but converge to finite values when $J_{F}$$%
\rightarrow\infty$.

\section{Valence Bond Ground State for the Haldane-like phase}

\begin{figure}[tbp]
\includegraphics[width=1.0\linewidth]{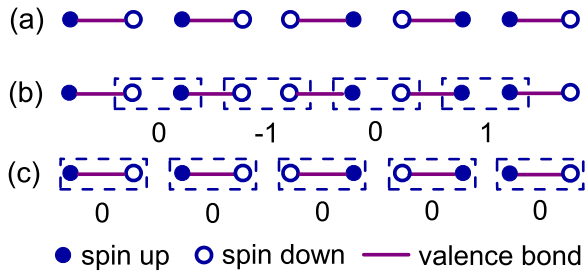}
\caption{(Color online) A schematic valence bond picture for the gapped ground state
of the tetrameric Heisenberg chain. (a) A typical configuration of the
valence bond state for the tetrameric chain in the absence of a magnetic
field. The solid bullet denotes spin up, the hollow bullet denotes spin
down, and the short line denotes the singlet bond. (b) The nonvanishing
string order. (c) The vanishing string order.}
\label{VBS}
\end{figure}

The nature of the spin-$1$ Heisenberg AF model can be demonstrated by the
AKLT model, where a Hamiltonian with the valence bond is constructed \cite%
{AKLT}. The Haldane-like gapped phase of this tetrameric system implies that
it is reasonable to expect a valence bond ground state that could explain
the numerical results and support our analysis. As discussed above, the gap
and string order are related to the singlet states of the spins coupled by $%
J_{AF_{1}}$. Therefore, a valence-bond-solid (VBS) state picture for the
ground state of the tetrameric chain can be proposed, as shown in Fig. \ref%
{VBS} (a). The singlet valence bonds, which are represented as short lines,
form between the spins coupled by $J_{AF_{1}}$. The gap, which is the energy
needed to break the bonds, should increase with $J_{AF_{1}}$. This has been
confirmed by our numerical results. After applying a magnetic field, the
magnetization plateau at $m=1/4$ could appear when the bonds are broken, and
the width of the plateau is thus mainly determined by $J_{AF_{2}}$, as
analyzed in Sec. \uppercase\expandafter{\romannumeral3}. In Figs. \ref{VBS}
(b) and (c), the exponential parts of the two topologically different string
orders are encircled by dashed lines. The numbers under the lines denote the
spin $(S_{i}^{z}+S_{i+1}^{z})$. The nonvanishing string order has the hidden
N\'{e}el order, as shown in Fig. \ref{VBS} (b). But in Fig. \ref{VBS} (c),
the vanishing string order does not exhibit the hidden N\'{e}el order in
this picture.

In the absence of a magnetic field, we propose a trial wavefunction $|\Omega
\rangle $, which is defined as a linear combination of two functions $|\Psi
\rangle $ and $|\Phi \rangle $, i.e. $|\Omega \rangle $=A$|\Psi \rangle $+B$%
|\Phi \rangle $, for this VBS state picture. The wavefunction $|\Psi \rangle 
$ is defined as 
\begin{equation}
|\Psi \rangle =\psi _{\alpha _{1}}\otimes \psi _{\alpha _{2}}\otimes \cdots
\otimes \psi _{\alpha _{4L-1}}\otimes \psi _{\alpha _{4L}}\varepsilon
^{\alpha _{1}\alpha _{2}}\cdots \varepsilon ^{\alpha _{4L-1}\alpha _{4L}},
\end{equation}%
where $\psi _{\alpha _{i}}$ ($\alpha _{i}=1,2$) denote the eigenstates of $%
S^{z}$ with eigenvalues $1/2$ and $-1/2$ of the spin on the site $i$ and $%
\varepsilon ^{\alpha \beta }$ is the antisymmetric tensor with $\varepsilon
^{12}=1/\sqrt{2}$. These wavefunctions are written under periodic boundary
conditions and thus $L$ is even. This function is one of the ground states
of the Majumdar-Ghosh model and qualitatively measures the ground states of
this tetrameric system in large $J_{AF_{1}}$ limit. In the state $|\Psi
\rangle $, the system has the maxima for both the gap and the string order.
The singlet correlation of the spins coupled by $J_{AF_{1}}$ also has the
maximum. In order to give a trial function that could depict the ground
state, we introduce another wavefunction $|\Phi \rangle $, which is defined
as 
\begin{eqnarray}
|\Phi \rangle &=&\psi _{\alpha _{2}}\otimes \psi _{\alpha _{3}}\varepsilon
^{\alpha _{2}\alpha _{3}}\otimes \psi _{\alpha _{4}}\otimes \psi _{\alpha
_{5}}\lambda ^{\alpha _{4}\alpha _{5}}\otimes \cdots \otimes \psi _{\alpha
_{4L-2}}  \nonumber \\
&\otimes &\psi _{\alpha _{4L-1}}\varepsilon ^{\alpha _{4L-2}\alpha
_{4L-1}}\otimes \psi _{\alpha _{4L}}\otimes \psi _{\alpha _{1}}\lambda
^{\alpha _{4L}\alpha _{1}},
\end{eqnarray}%
where $\lambda ^{\alpha \beta }$ is the symmetric tensor with $\lambda
^{11}=\lambda ^{22}=0$ and $\lambda ^{12}=1/\sqrt{2}$. $|\Phi \rangle $
measures the states in both large $J_{AF_{2}}$ and $J_{F}$ limit. In this
state, the system is gapless and the string order is vanishing.

In the thermodynamic limit $L\rightarrow \infty$, $\vert \Psi \rangle$ and $%
\vert \Phi \rangle$ are orthogonal, namely $\langle \Psi\vert\Phi \rangle =0$%
. The normalization of the trial function $\vert \Omega \rangle$ thus
requires $A^{2}+B^{2}=1$. We have performed calculations of the physical
properties on this trial function in the thermodynamic limit. It is found
that the trial function $\vert \Omega \rangle$ has vanishing local magnetic
moment $\langle S^{z}_{j}\rangle$ and a short-range spin correlation
function. The string order is found to be $O^{z}_{string}=\frac{1}{4}A^{2}$,
which is finite in the state $\vert \Omega \rangle$ and could well fit into
the numerical results shown in Figs. \ref{String} (a) and (b) when the
coefficient $A$ is chosen as 
\begin{eqnarray}
A^{2}&=&(\frac{4-J_{AF_{2}}/J_{AF_{1}}}{3\sqrt{4-2J_{AF_{2}}/
J_{AF_{1}}+(J_{AF_{2}}/J_{AF_{1}})^{2}}}+\frac{1}{3})  \nonumber \\
&\times&(\frac{4+J_{F}/J_{AF_{1}}}{3\sqrt{4+2J_{F}/
J_{AF_{1}}+(J_{F}/J_{AF_{1}})^{2}}}+\frac{1}{3}).
\end{eqnarray}
The numerical results of the gap as well as this valence-bond-state picture
indicate that the gap is induced by the valence bond, namely the singlet
state between the spins coupled by $J_{AF_{1}}$. The string order could well
describe the gap and the singlet correlation that measures the singlet state
of the spins coupled by $J_{AF_{1}}$. This above picture shows the crucial
role of the valence bond in the formation of the Haldane-like gap and hidden
symmetry.

\section{Summary and Discussion}

By means of the DMRG method, we have studied the magnetic properties, spin
gap, spin-spin correlation function, and string order of the spin-$1/2$
tetrameric HAFC with alternating couplings AF$_{1}$-AF$_{2}$-AF$_{1}$-F. Two
magnetization plateaux at $m=0$ and $1/4$ are found, which satisfy the OYA
condition for such a spin chain with a period of $4$. For the $m=1/4$
plateau states, an approximate wave function is proposed, which fits into
the numerical results of the local magnetic moment and spin correlation
function perfectly. The magnetization process of the two regions between the
plateaux is fitted well by Eq. (\ref{fit}).

Besides the $m=1/4$ plateau, the gapped ground state in the absence of the
magnetic field is also investigated. The system is found to have a gap from
the singlet ground state to the triplet excited states. The gap changes
monotonically with the couplings and does not show any level-crossing. It
decreases with $J_{F}$ and $J_{AF_{2}}$, but increases with $J_{AF_{1}}$. 
The ground state energy has no nonanalyticity in the parameter space, 
which implies an absence of the quantum phase transition. The asymptotic
behaviors of the gap have been calculated, which have the same features as
those in the isolated tetramer systems. Combining the numerical results with
the nonlinear $\sigma$ model, it is shown that a gap would be generated by
an arbitrary small $J_{AF_{1}}$ and thus this system is always gapped if $%
J_{AF_{1}}\neq 0$.

The spin-spin correlation function is uncovered to decay exponentially in
the gapped states, but the correlation length and the behaviors of the
correlation function show dramatic changes for different couplings, which
are observed in the variations of the peaks of the static structure factor $%
S(q)$, reflecting a complex competition of the interactions. When $%
J_{AF_{2}} $ or $J_{F}$ is small compared with $J_{AF_{1}}$, $S(q)$ is
dominated by short-range correlations. It would exhibit three maxima at $%
q=\pi/3 ,\pi$ and $5\pi /3$ when $J_{F}$ is small. When $J_{AF_{2}}$ is
small, the maximum at $q=\pi$ becomes a valley and two small maxima appear
near $q=3\pi /5$ and $7\pi /5$. In contrast, when both $J_{F}$ and $%
J_{AF_{2}}$ are large enough, $S(q)$ exhibits four peaks at $q=\pi /4,3\pi
/4,5\pi /4$ and $7\pi /4$, which is due to the increase of the correlation
length and the translation symmetry of the ground states.

In order to investigate the hidden symmetry of the gapped ground states, we
proposed a string order and performed calculations to detect the hidden
symmetry. The string order is found to be nonvanishing in the gapped phase,
indicating a hidden symmetry. More importantly, the behaviors of the string
order have the same tendencies as that of the gap. The opening of the gap
and its variations could be well described by this nonlocal order. The gap
is related to the singlet state of the spins coupled by $J_{AF_{1}}$ and
increases with this interaction. Thus, the gap could be regarded to be
determined by the singlet correlation. The nonlocal string order could also
measure this local singlet correlation. The singlet state of the spins
coupled by $J_{AF_{1}}$ provides a picture to understand the hidden N\'{e}el
order reflected by the finite string order. Different from the conventional
hidden N\'{e}el order, the order in the present system is not of the singlet
spins but of the spin pairs coupled by $J_{AF_{2}}$ or $J_{F}$.

The hidden symmetry measured by the string order is unveiled by a dual
transformation. The present tetrameric Hamiltonian is transformed into a
quantum AT model, which has a discrete $Z_{2}\times Z_{2}$ symmetry. The
string order becomes a ferromagnetic order that is proper to measure this
discrete symmetry. The nonvanishing string order indicates that this $%
Z_{2}\times Z_{2}$ symmetry is fully breaking in the gapped phase, which is
the characteristic of Haldane phase. Thus, this tetrameric chain is in a
Haldane-like gapped phase.

A valence-bond-solid state picture and a trial function for the gapped
states are proposed, most properties of which could be explained on the
basis of this picture. The numerical results as well as the VBS picture
indicate that the gap is induced by the valence bond of the spins coupled by 
$J_{AF_{1}}$. The valence bond is measured by the singlet correlation. The
string order could well describe not only the gap, but also the singlet
correlation. The critical role of the valence bond on the formation of the
Haldane-like gap and hidden symmetry is unveiled.

Finally, we would like to state that unlike the spin-$1/2$ F-AF alternating
HAFC, the present spin-$1/2$ tetrameric HAFC could not reduce to an
integer-spin HAFC. Therefore, our findings extend the substance of Haldane's
conjecture that was originally proposed for HAFCs with integer spin. We
expect that our investigations would deepen further understanding on the
physical properties of low dimensional quantum magnetism.

\acknowledgments

We are grateful to X. Chen, W. Li, X. L. Sheng, Z. C. Wang, Z. Xu, Q. B.
Yan, L. Z. Zhang, Q. R. Zheng and G. Q. Zhong for useful discussions. This
work is supported in part by the National Science Fund for Distinguished
Young Scholars of China (Grant No. 10625419), the National Science
Foundation of China (Grant Nos. 90403036 and 20490210), the MOST of
China (Grant No. 2006CB601102), and the Chinese Academy of Sciences.


\begin{thebibliography}{99}
\bibitem[]{Permanent address.} $^{\ast }$Corresponding author. E-mail:
gsu@gucas.ac.cn

\bibitem{Haldane} F. D. Haldane, \textit{Phys. Rev. Lett.} \textbf{50}, 1153
(1983); \textit{Phys. Lett.} \textbf{93A}, 464 (1983).

\bibitem{AKLT} I. Affleck, T. Kennedy, E.H. Lieb, and H. Tasaki, \textit{%
Phys. Rev. Lett.} \textbf{59}, 799 (1987); \textit{Commun. Math. Phys.} 
\textbf{115}, 477 (1988).

\bibitem{NR} M.P.M. den Nijs and K. Rommelse, \textit{Phys. Rev. B} \textbf{%
40}, 4709 (1989).

\bibitem{KT} Tom Kennedy and Hal Tasaki, \textit{Phys. Rev. B} \textbf{45},
304 (1992).

\bibitem{Hida1} K. Hida, \textit{J. Phys. Soc. Jpn.} \textbf{60}, 1347
(1991).

\bibitem{WNT} H. Watanabe, K. Nomura, and S. Takada, \textit{J. Phys. Soc.
Jpn.} \textbf{62}, 2845 (1993).

\bibitem{BR} T. Barnes and J. Riera, \textit{Phys. Rev. B} \textbf{50}, 6817
(1994).

\bibitem{Hida2} K. Hida, \textit{Phys. Rev. B} \textbf{45}, 2207 (1992).

\bibitem{Gu} B. Gu, G. Su, and S. Gao, \textit{J. Phys.: Condens. Matter} 
\textbf{17}, 6081 (2005); B. Gu, G. Su, and S. Gao, \textit{Phys. Rev. B} 
\textbf{73}, 134427 (2006).

\bibitem{HMK} M. Hagiwara, K. Minami, and H. A. Katori, \textit{Prog. Theor.
Phys. Suppl.} \textbf{145}, 150 (2002).

\bibitem{HNMK} M. Hagiwara, Y. Narumi, K. Minami, K. Kindo, H. Kitazawa, H.
Suzuki, N. Tsujii, and H. Abe, \textit{J. Phys. Soc. Jpn.} \textbf{72}, 943
(2003).

\bibitem{EVF} A. Escuer, R. Vicente, M. S. El Fallah, M. A. S. Goher, and F.
A. Mautner, \textit{Inorg. Chem.} \textbf{37}, 4466 (1998).

\bibitem{Yamamoto} S. Yamamoto, \textit{Phys. Rev. B} \textbf{69}, 064426
(2004).

\bibitem{NY} T. Nakanishi and S. Yamamoto, \textit{Phys. Rev. B} \textbf{65}%
, 214418 (2002).

\bibitem{LSS} H. T. Lu, Y. H. Su, L. Q. Sun, J. Chang, C. S. Liu, H. G. Luo,
and T. Xiang, \textit{Phys. Rev. B} \textbf{71}, 144426 (2005).

\bibitem{37Mn} M. Yuan, Z. M. Wang, and S. Gao, unpublished.

\bibitem{DMRG1} S. R. White, \textit{Phys. Rev. Lett.} \textbf{69}, 2863
(1992); S. R. White, \textit{Phys. Rev. B} \textbf{48}, 10345 (1993).

\bibitem{DMRG2} U. Schollw\"{o}ck, \textit{Rev. Mod. Phys.} \textbf{77}, 259
(2005).

\bibitem{Hida} K. Hida, \textit{J. Phys. Soc. Jpn.} \textbf{63}, 2359 (1994).

\bibitem{Okamoto} K. Okamoto, \textit{Solid State Commun.} \textbf{98}, 245
(1996).

\bibitem{OYA} M. Oshikawa, M. Yamanaka, and I. Affleck, \textit{Phys. Rev.
Lett.} \textbf{78}, 1984 (1997).

\bibitem{Gu2} B. Gu and G. Su, \textit{Phys. Rev. Lett.} \textbf{97}, 089701
(2006); \textit{Phys. Rev. B} \textbf{75}, 174437 (2007).

\bibitem{Cabra} D. C. Cabra and M. D. Grynberg, \textit{Phys. Rev. B} 
\textbf{59}, 119 (1999).

\bibitem{AL} I. Affleck and E. H. Lieb, \textit{Lett. Math. Phys.} \textbf{12%
}, 57 (1986).

\bibitem{LSM} E. H. Lieb, T. D. Schultz, and D. C. Mattis, \textit{Ann. Phys.%
} \textbf{16}, 407 (1961).

\bibitem{QFT} S. Sachdev, \textit{Quantum Phase Transitions} (Cambridge
University Press , Cambridge, 1999).

\bibitem{NLSM1} I. Affleck, \textit{Nucl. Phys. B} \textbf{257}, 397 (1985); 
\textbf{265}, 409 (1986); I. Affleck and F. D. M. Haldane, \textit{Phys.
Rev. B} \textbf{36}, 5291 (1987).

\bibitem{NLSM2} T. Fukui and N. Kawakami, \textit{Phys. Rev. B} \textbf{56},
8799 (1997).

\bibitem{NLSM3} K. Takano, \textit{Phys. Rev. Lett.} \textbf{82}, 5124
(1999).

\bibitem{FAF} We have confirmed this result of the static structure
factor S(q) for the spin-1/2 alternating F-AF Heisenberg
chain by means of the DMRG method.

\bibitem{GSS} S. S. Gong, B. Gu, and G. Su, \textit{Phys. Lett. A} \textbf{%
372}, 2322 (2008).

\bibitem{YHK} M. Yamanaka, Y. Hatsugai, and M. Kohmoto, \textit{Phys. Rev. B}
\textbf{48}, 9555 (1993).

\bibitem{SV} E. H. Kim, G. F\'{a}th, J. S\'{o}lyom, and D. J. Scalapino, 
\textit{Phys. Rev. B} \textbf{62}, 14965 (2000).

\bibitem{SNT} D. G. Shelton, A. A. Nersesyan, and A. M. Tsvelik, \textit{%
Phys. Rev. B} \textbf{53}, 8521 (1996).

\bibitem{MH} M. Kohmoto and H. Tasaki, \textit{Phys. Rev. B} \textbf{46},
3486 (1992).

\bibitem{KNK} M. Kohmoto, M. den Nijs, and L. P. Kadanoff, \textit{Phys.
Rev. B} \textbf{24}, 5229 (1981).

\bibitem{KW} H. A. Kramers and G. H. Wannier, \textit{Phys. Rev.} \textbf{60}%
, 252 (1941).
\end{thebibliography}
\end{document}